\documentclass[%
reprint,
superscriptaddress,
 amsmath,amssymb,
 prl,
]{revtex4-2}

\usepackage{graphicx}
\usepackage{dcolumn}
\usepackage{bm}

\begin{document}

\preprint{APS/123-QED}

\title{Observations of Electromagnetic Electron Holes and Evidence of Cherenkov Whistler Emission}


\author{Konrad Steinvall}
\email{konrad.steinvall@irfu.se}
\affiliation{Swedish Institute of Space Physics, Uppsala, 75121, Sweden}
\affiliation{Space and Plasma Physics, Department of Physics and Astronomy, Uppsala University, Uppsala, 75120, Sweden.}

\author{Yuri V. Khotyaintsev}
\author{Daniel B. Graham}
\affiliation{Swedish Institute of Space Physics, Uppsala, 75121, Sweden}

\author{Andris Vaivads}
\affiliation{%
Division of Space and Plasma Physics, School of Electrical Engineering and Computer Science, KTH Royal Institute of Technology, Stockholm, 11428, Sweden
}%

\author{Olivier Le Contel}
\affiliation{%
Laboratoire de Physique des Plasmas, CNRS/Ecole Polytechnique/Sorbonne Universit\'e/Univ. Paris Sud/Obs. de Paris, Paris, F-75252 Paris Cedex 05, France
}%

\author{Christopher T. Russell}
\affiliation{%
Department of Earth and Space Sciences, University of California, Los Angeles, California, 90095, USA
}%

\date{\today}

\begin{abstract}

We report observations of electromagnetic electron holes (EHs). We use multi-spacecraft analysis to quantify the magnetic field contributions of three mechanisms: the Lorentz transform, electron drift within the EH, and Cherenkov emission of whistler waves. The first two mechanisms account for the observed magnetic fields for slower EHs, while for EHs with speeds approaching half the electron Alfv\'en speed, whistler waves excited via the Cherenkov mechanism dominate the perpendicular magnetic field. The excited whistlers are kinetically damped and typically confined within the EHs.

\end{abstract}

\maketitle

Electron holes (EHs) are localized nonlinear plasma structures in which electrons are self-consistently trapped by a positive potential \cite{schamel1979,schamel1972,hutchinson2017}. By scattering and heating electrons, EHs play an important part in plasma dynamics \cite{che2010,vasko2017}. EHs are frequently observed in space \cite{matsumoto1994,ergun1998,pickett2008,norgren2015,graham2016} and laboratory \cite{lynov1979,fox2008,lefebvre2010} plasmas. They are typically manifested in data as diverging, bipolar, electric fields parallel to the ambient magnetic field.
EHs are formed by various instabilities \cite{omura1996,miyake1998}, and are thus indicators of prior instability and turbulence. Their connection with streaming instabilities leads them to frequently appear during magnetic reconnection \cite{drake2003,cattell2005,khotyaintsev2010,divin2012}. Furthermore, simulations of magnetic reconnection have shown EHs can Cherenkov radiate whistler waves which in turn affect the reconnection rate \cite{goldman2014}. Studying EHs can thus prove important for understanding key plasma phenomena such as magnetic reconnection.

Though EHs are usually considered electrostatic, observations of electromagnetic EHs have been made in Earth's magnetotail \cite{andersson2009,lecontel2017}. The observed magnetic fields ($\delta\mathbf{B}$) were argued to be the sum of two independent fields. First, $\delta\mathbf{B}_L$ generated by the Lorentz transform, of the electrostatic field, and second, $\delta\mathbf{B}_d$ generated by the $\delta\mathbf{E}\times\mathbf{B}_0$ drift of electrons associated with the EH electric field and ambient magnetic field \cite{andersson2009,tao2011}. These studies were limited either by the fact that the EHs were only observed at one point in space \cite{andersson2009}, or provided only estimates of $\delta B_{d\parallel}$ at the EH center \cite{lecontel2017}. With the Magnetospheric Multiscale (MMS)~\cite{burch2016} mission, it is possible to use four-spacecraft measurements to obtain a complete three-dimensional description of EHs \cite{holmes2018,tong2018,steinvall2018}, enabling $\delta\mathbf{B}$ to be investigated in greater detail \cite{holmes2018}.

In this letter we use data from MMS to investigate electromagnetic EHs frequently observed during boundary layer crossings in the magnetotail. We use multi-spacecraft methods to quantify different contributions to $\delta\mathbf{B}$. Our results show that $\delta B_{d,\parallel}$ well explains the observed $\delta B_\parallel$, and that $\delta\mathbf{B}_{d,\perp}$ is in good agreement with observations for EHs that are much slower than the electron Alfv\'en speed. For increasing EH speeds we show, for the first time, that localized whistler waves are excited from the EHs via the Cherenkov mechanism and contribute significantly to $\delta\mathbf{B}_\perp$.

\begin{figure*}
\includegraphics[width=1\linewidth]{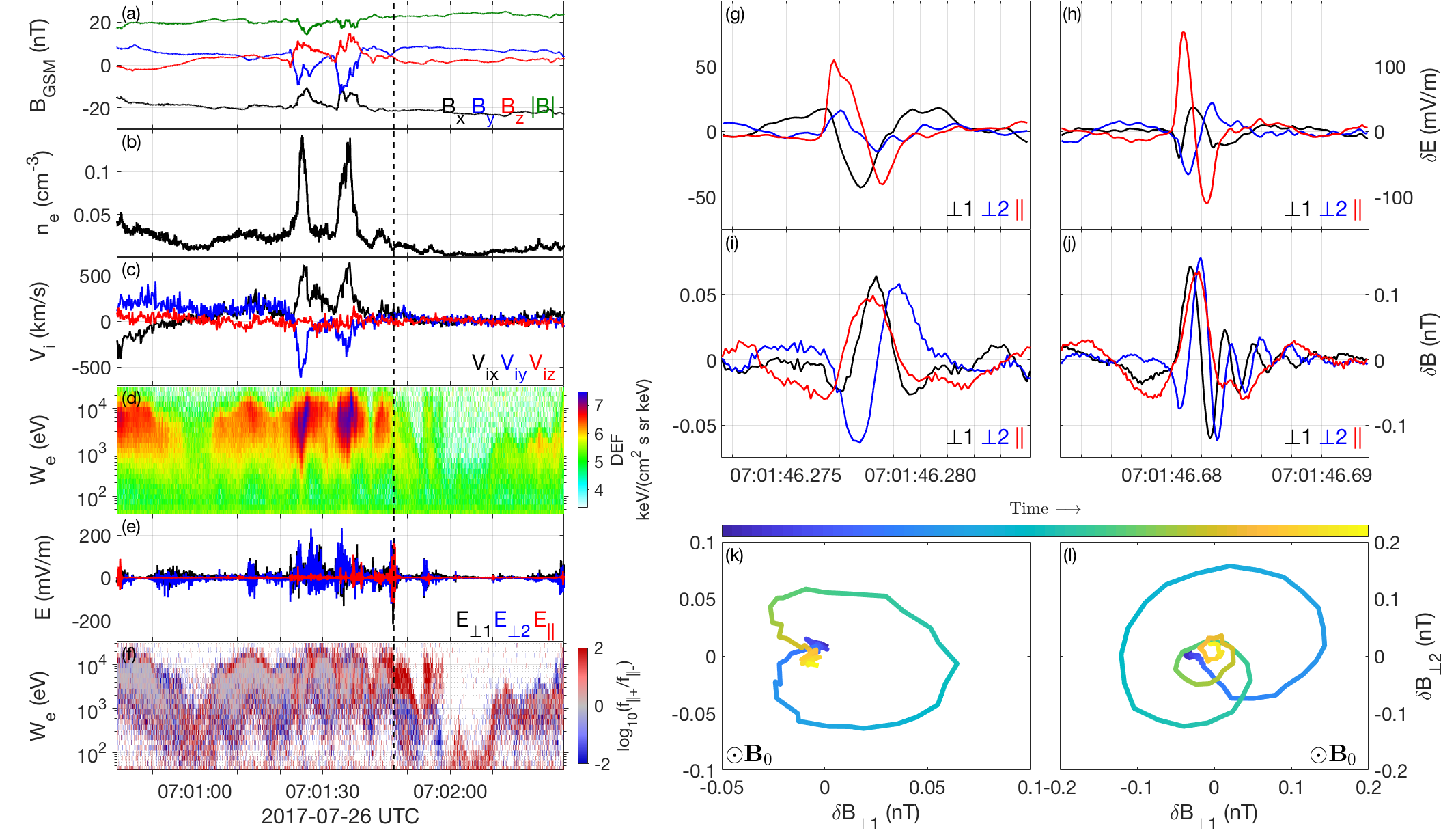}
\caption{\label{fig:overview} Left: Event overview. (a) Magnetic field from FGM~\cite{russell2016} in geocentric solar magnetospheric (GSM) coordinates, (b) plasma density from FPI~\cite{pollock2016}, (c) ion velocity from FPI in GSM, (d) electron energy spectrogram from FPI, (e) electric field from EDP~\cite{lindqvist2016,ergun2016} in field-aligned coordinates, (f) spectrogram of the ratio of the parallel and anti-parallel electron phase-space density from FPI. The vertical dashed line shows where EHs are observed. Right: Examples of electromagnetic EHs. The data is high-pass filtered at 100 Hz. (g,h) Electric field from EDP, (i,j) magnetic field from SCM~\cite{lecontel2016}, (k,l) hodograms of $\delta\mathbf{B}_\perp$. }
\end{figure*}

Fig.~\ref{fig:overview} shows an example of a plasma sheet boundary layer crossing containing signatures of magnetic reconnection and EHs with magnetic fields.
At 2017-07-26 07:00\,UT, MMS was in the plasma sheet and detected a fast reconnection jet moving tailward (Fig.~\ref{fig:overview}c). At 07:01:30, the ion flow reversed, and MMS entered the boundary layer between the plasma sheet and the tail lobes (Fig.~\ref{fig:overview}d) where strong wave activity was observed (Fig.~\ref{fig:overview}e). First as low-frequency $E_\perp$ oscillations consistent with lower hybrid drift waves~\cite{norgren2012}, and later as solitary $E_\parallel$ waves marked by the vertical dashed line in Fig.~\ref{fig:overview}e, and exemplified in Figs.~\ref{fig:overview}g,h. The solitary waves were accompanied by a high-energy electron beam (Fig.~\ref{fig:overview}f) parallel to $\mathbf{B}_0$. By timing $E_\parallel$ between the spacecraft we find the structures to be EHs moving together with the beam. Notably the EHs have magnetic field fluctuations $\delta\mathbf{B}$ associated with them. We show two EH examples in Figs.~\ref{fig:overview}g-j. While both EHs have positive and monopolar $\delta B_\parallel$ (distorted in the figure by high-pass filtering) confined within the EH, there are significant differences in $\delta\mathbf{B}_\perp$. For the first EH (Figs.~\ref{fig:overview}g,i), $\delta\mathbf{B}_\perp$ is localized within the EH, whereas for the second EH, $\delta\mathbf{B}_\perp$ oscillates multiple times and forms a trailing tail (Fig.~\ref{fig:overview}h,j). Note that of the roughly 40 EHs that were observed during this time, only two EHs had the tail-like feature in Fig.~\ref{fig:overview}j, the others resembled Fig.~\ref{fig:overview}i. The polarization of $\delta\mathbf{B}_\perp$ is right handed for all cases (Figs.~\ref{fig:overview}k,l) with dominant frequency $\omega\approx0.7\Omega_{ce}<\omega_{pe}$, where $\Omega_{ce}$ and $\omega_{pe}$ are the electron cyclotron and plasma frequencies.

We perform a statistical study to investigate how $\delta\mathbf{B}$ depends on EH properties. To accurately estimate the electron hole speed, $v_{EH}$, and  parallel length scale, $l_\parallel$, the EHs should be detected by as many spacecraft as possible, and all four spacecraft are needed to accurately estimate the EH center potential, $\Phi_0$, and perpendicular length scale, $l_\perp$~\cite{tong2018,steinvall2018}. We therefore limit the study to June-August 2017, when MMS was probing the magnetotail with electron scale spacecraft separation. We take 9 data intervals where one or more groups of electromagnetic EHs are observed, resulting in a data-set of 336 EHs, all observed in connection to boundary layers similar to that in Fig.~\ref{fig:overview}. 
  
\begin{figure}
\includegraphics[width=1\linewidth]{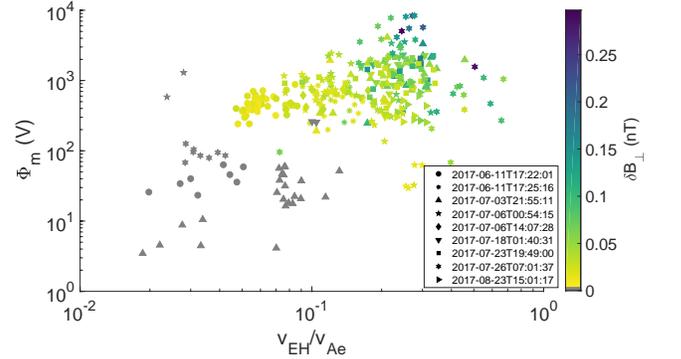}
\caption{\label{fig:stats} Measured EH potential $\Phi_m$ against $v_{EH}/v_{Ae}$ for 336 EHs, with the peak value of $\delta B_\perp$ color-coded. EHs from the same burst-data interval have the same symbol.}
\end{figure}

We use the multi-spacecraft timing method discussed in Ref.~\onlinecite{steinvall2018}, cross-correlating $\delta E_\parallel$ between the spacecraft, to determine $\bm{v}_{EH}$, $l_\parallel$, and the measured potential $\Phi_m=\int{\delta E_\parallel v_{EH}dt}$ of the 336 EHs. The median propagation angle of the EHs with respect to $\mathbf{B}_0$ is 12$^\circ$ which is within the uncertainty of the four-spacecraft timing, so $\bm{v}_{EH}$ is assumed to be field aligned. In Fig.~\ref{fig:stats} we plot $\Phi_m$ against $v_{EH}/v_{Ae}$ ($v_{Ae}=c\Omega_{ce}/\omega_{pe}$ is the electron Alfv\'en speed), with the peak value of $\delta\mathbf{B}_\perp$ color-coded. The figure shows that $\delta\mathbf{B}_\perp$ increases with potential and velocity. A dependence on $\Phi_m$ is expected since $\delta\mathbf{B}_L,\delta\mathbf{B}_d\propto \delta \mathbf{E}_\perp\propto\Phi_0$ and the $v_{EH}/v_{Ae}$ dependence is qualitatively consistent with $\delta\mathbf{B}_L\propto v_{EH}$ since the EHs were observed in the same plasma region with, for the most part, similar $v_{Ae}$.

\begin{figure}
\includegraphics[width=1\linewidth]{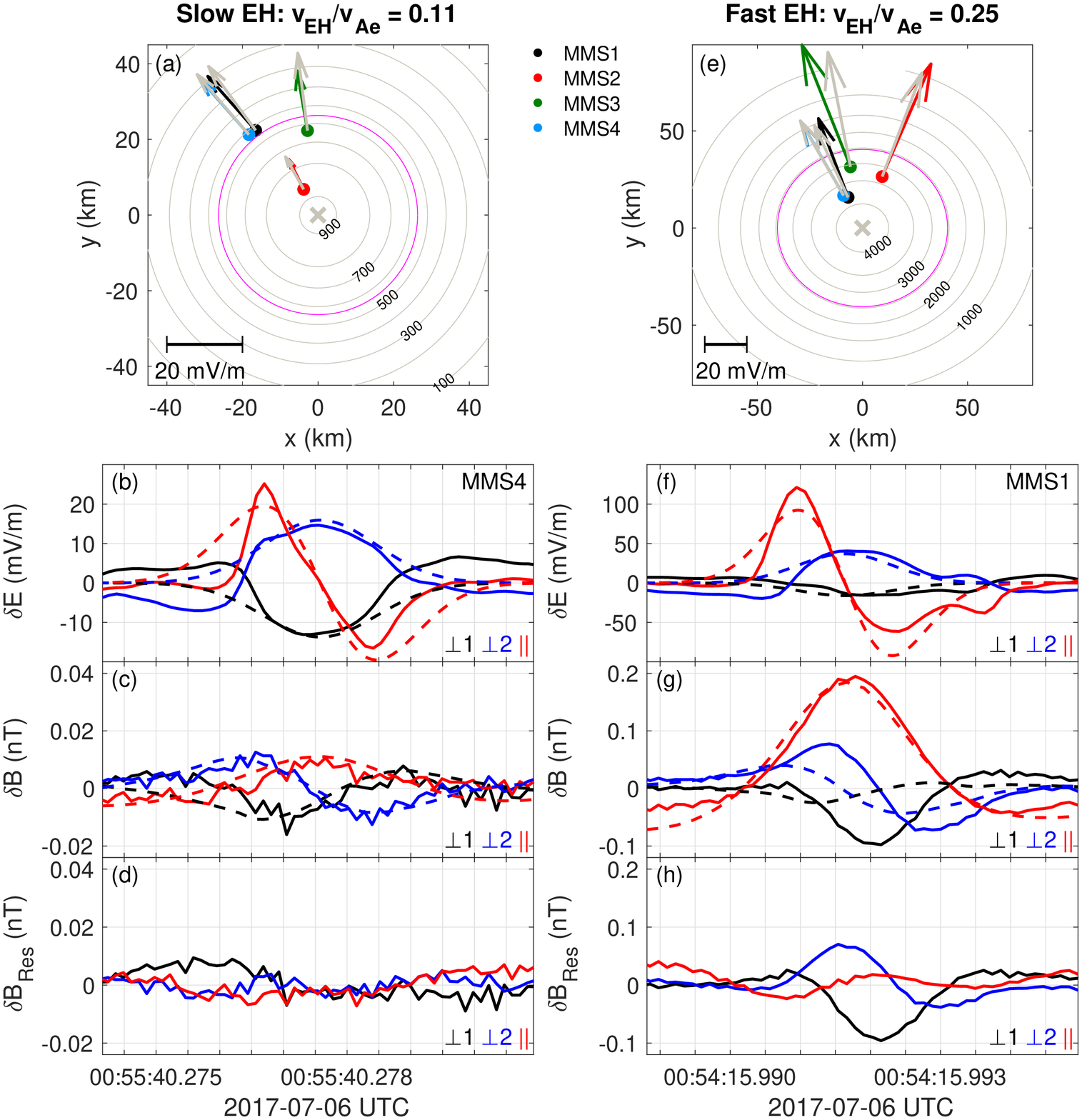}
\caption{\label{fig:fit} Two examples of EH fits and induced magnetic fields. (a) The position of MMS (colored dots) and the EH (grey cross) in the perpendicular plane. The measured and fitted $\delta\mathbf{E}_\perp$ are illustrated by the colored and grey arrows, respectively, where the arrow length is proportional to $|\delta\mathbf{E}_\perp|$. The grey contours are EH equipotential lines in Volts, and the magenta circle corresponds to $r=l_\perp$. (b) Measured (solid) and fitted (dashed) $\delta\mathbf{E}$. (c) Measured $\delta\mathbf{B}$ (solid) and calculated $\delta\mathbf{B}_L+\delta\mathbf{B}_d$ (dashed). (d) $\delta\mathbf{B}-\delta\mathbf{B}_L-\delta\mathbf{B}_d$. (e)-(h) Same format as (a)-(d) for a different EH. All fields are high-pass filtered at 50\,Hz.}
\end{figure}

Next, we investigate the different mechanisms that can generate $\delta\mathbf{B}$. 
For weakly relativistic EHs (i.e. $\gamma\approx1$)
$\delta B_{L,\{\perp1,\perp2\}}=\mp v_{EH}\delta E_{\{\perp2,\perp1\}}/c^2$ \cite{jackson1999}. By assuming the EH potential
\begin{equation}
\label{eq:pot}
    \Phi(r,\theta,z)=\Phi_0 e^{-r^2/2l_\perp^2}e^{-z^2/2l_\parallel^2},
\end{equation}
$\delta\mathbf{B}_d$ is given by the Biot-Savart law of the $\delta\mathbf{E}\times\mathbf{B}_0$ current $J_\theta=e n_0r\Phi(r,z)/(B_0l_\perp^2)$ \cite{tao2011} as 
\begin{equation}
\label{eq:biot}
    \delta\mathbf{B}_d(\bm{x}) = \frac{en_0\mu_0}{4\pi B_0}\int{\frac{r'}{l_\perp^2}\Phi(r',z')\hat{\bm{\theta}}\times\frac{\bm{x}-\bm{x}'}{|\bm{x}-\bm{x}'|^3}d^3x'},
\end{equation}
where $n_0$ is the electron density, and $e$ is the elementary charge.
In Fig.~\ref{fig:fit} we show two examples of EHs where we calculate and compare $\delta\mathbf{B}_L$ and $\delta\mathbf{B}_d$ with observations. The first EH (Figs.~\ref{fig:fit}a-d) is small amplitude ($\Phi_m = 680$\,V), slow ($v_{EH}/v_{Ae}= 1/9$) and has a weak $\delta\mathbf{B}\sim 0.01$ nT. We use the method of Ref.~\onlinecite{tong2018} (using, instead of the maximum value, $\delta\mathbf{E}_\perp$ evaluated at $\delta E_\parallel = 0$) to fit the $\delta\mathbf{E}$ data of the four spacecraft to the electrostatic field corresponding to Eq.~(\ref{eq:pot}), giving $l_\perp=26$ km $=0.6d_e=1.6l_\parallel$, where $d_e=c/\omega_{pe}$ is the electron inertial length; $\Phi_0=915$\,V $=1.4 T_e/e$, where $T_e$ is the electron temperature; and the position of the EH. A representation of the fit is shown in Fig.~\ref{fig:fit}a, where we plot the spacecraft (colored dots) and the EH (grey cross) position in the perpendicular plane. The arrows are the measured (colored) and predicted (grey) $\delta\mathbf{E}_\perp$ evaluated at $\delta E_\parallel=0$, showing that the EH fit well describes $\delta\mathbf{E}_\perp$ for all four spacecraft.
A time series representation of the fit is shown in Fig.~\ref{fig:fit}b for MMS4, where the measured and fitted $\delta\mathbf{E}$ are the solid and dashed lines respectively, affirming that the fit is in good agreement with observations.
With $\Phi_0$, $l_\parallel$ and $l_\perp$ known, we solve Eq.~(\ref{eq:biot}) numerically to obtain $\delta\mathbf{B}_d$. $\delta\mathbf{B}_L$ is small,  $|\delta\mathbf{B}_L|\approx0.004$\,nT. We plot MMS4 data of $\delta\mathbf{B}$ (solid) together with $\delta\mathbf{B}_L+\delta\mathbf{B}_d$ (dashed) in Fig.~\ref{fig:fit}c, and the residual $\delta\mathbf{B}_\text{Res}=\delta\mathbf{B}-\delta\mathbf{B}_L-\delta\mathbf{B}_d$ in Fig.~\ref{fig:fit}d. We find that $\delta\mathbf{B}\approx\delta\mathbf{B}_d$, the only discrepancy being that $|\delta B_{d,\perp1}|$ is overestimated initially. This might be due to the fact that the EH has a steeper increase of $\delta E_\parallel$ than the model (Fig.~\ref{fig:fit}b).
The second EH (Fig.~\ref{fig:fit}e-h) has larger amplitude ($\Phi_m = 3.5$ kV), is faster ($v_{EH}/v_{Ae}= 1/4$) and has a stronger $\delta\mathbf{B}\sim 0.1$ nT. We perform the same analysis and present analogous plots in Fig.~\ref{fig:fit}e-h. As before, the EH fit of $\delta\mathbf{E}$ (Fig.~\ref{fig:fit}e,f) agrees well with observations ($\Phi_0=4.2$\,kV $=1.9 T_e/e$ and $l_\perp=40$\,km $=1.1d_e=1.6l_\parallel$), $|\delta\mathbf{B}_L|\approx0.02$\,nT is small compared to $|\delta\mathbf{B}_{\perp}|$, and $\delta{B}_\parallel$ is well traced by $\delta B_{d,\parallel}$. However, when it comes to $\delta\mathbf{B}_\perp$ there is significant $\delta\mathbf{B}_{\text{Res},\perp}$ implying an additional mechanism is contributing to $\delta\mathbf{B}_\perp$. We note that $\delta\mathbf{B}_{\text{Res},\perp}$ is right hand polarized and its dominant frequency $f\approx400$ Hz is below $f_{ce}\approx 650$ Hz. We estimate the wave normal angle of $\delta\mathbf{B}_{\text{Res},\perp}$ by $k_\parallel/k_\perp=\delta B_\perp/\delta B_\parallel=2.6$, corresponding to a wave normal angle $21^\circ$. We thus find that while $\delta\mathbf{B}$ of the slower EH can be fully explained by $\delta\mathbf{B}_d$, the faster EH has an additional $\delta\mathbf{B}_{\text{Res},\perp}$ with features consistent with whistler waves.

We are able to apply this method and calculate $\delta\mathbf{B}_d$ for a total of 19 EHs. The remaining EHs were either not observed by all four spacecraft ($\sim$50$\%$), had $\delta\mathbf{E}$ that was qualitatively inconsistent with the assumed potential model, e.g. bipolar $\delta\mathbf{E}_\perp$ ($\sim$25$\%$), or gave fitting results deemed too different from observations to be useful ($\sim$15$\%$).
For these 19 EHs, $\delta B_\parallel$ is consistently well described by $\delta B_{d,\parallel}$, and $|\delta\mathbf{B}_L|\ll|\delta\mathbf{B}_{d,\perp}|$, meaning $\delta\mathbf{B}_d$ is more important for generating $\delta\mathbf{B}$ in the observed parameter range of Fig.~\ref{fig:stats}. For all 19 EHs, when $\delta\mathbf{B}_{\text{Res},\perp}\neq0$, it is right hand polarized with $\omega<\Omega_{ce}<\omega_{pe}$ which we interpret as being related to the whistler mode.

Because $\delta\mathbf{B}_{\text{Res},\perp}$ is localized to the EHs, we believe the EHs to be the source of the whistlers, rather than for example temperature anisotropy or Landau resonance. In fact, for most observations $T_{e\perp}/T_{e\parallel}<1$, so whistlers should not grow from temperature anisotropy. In this section we consider the generation of whistler waves from EHs via the Cherenkov mechanism, and show that this is consistent with our observations.

The theory of whistler waves Cherenkov emitted by EHs is developed and discussed in Ref.~\onlinecite{goldman2014}. In summary, the Cherenkov resonance condition is $\omega/k_\parallel=v_{EH}$ which specifies $\omega$ and $k_\parallel$ of the excited wave. Further, the ratio of the whistler electric field to that of the EH grows secularly (linearly in time) at a rate proportional to $(v_{EH}/v_{Ae})^4$, subject to $v_{EH}\leq v_{Ae}/2$.
 
\begin{figure}
\includegraphics[width=1\linewidth]{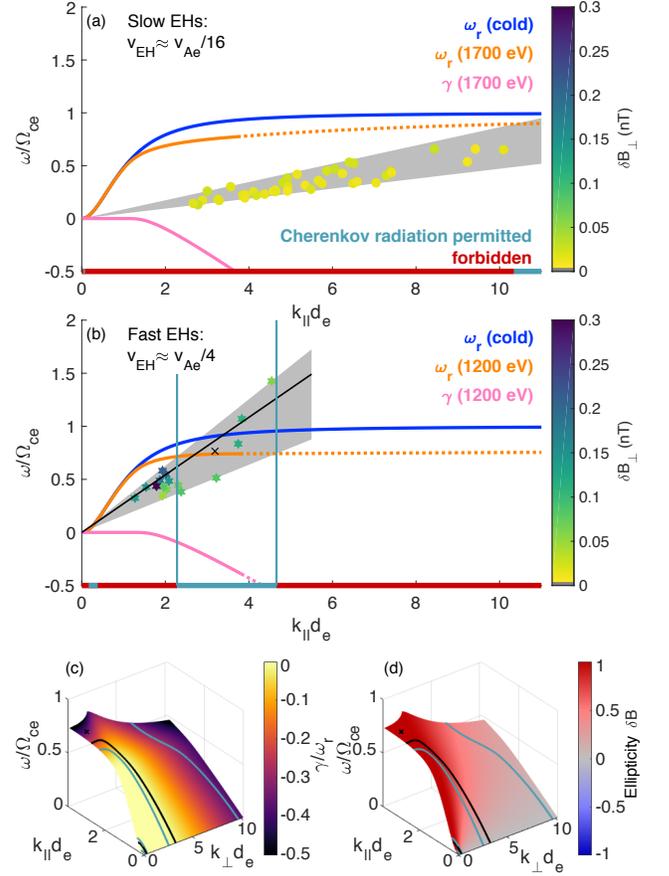}
\caption{\label{fig:cherenkov} (a)-(b) Cold (blue) and kinetic (orange and pink) whistler dispersion relation ($k_\perp=0$) for two different groups of EHs. The dotted lines are extrapolations (based on the cold plasma dispersion relation) of the kinetic results, and are not exact. EH data is plotted with symbols and colorbar consistent with Fig.~\ref{fig:stats}. The average $v_{EH}$ is $v_{Ae}/16$ in (a), and $v_{Ae}/4$ in (b). The shaded intervals show $\text{min}(v_{EH})\leq v\leq\text{max}(v_{EH})$, and the corresponding $k_\parallel$ intervals satisfying $\omega/k_\parallel=v_{EH}$ are marked in blue. The black line and cross in (b) show the EH speed and observed properties of $\delta\mathbf{B}_\perp$ in Fig.~\ref{fig:overview}j. (c) Whistler dispersion relation for $k_\perp\geq0$, color-coding the relative damping $\gamma/\omega_r$. The blue contours show the boundaries of the Cherenkov-permitted regions, and the black contour corresponds to the resonant waves of the EH in Fig.~\ref{fig:overview}h. (d) Same as (c), but with ellipticity of $\delta\mathbf{B}$ color coded, $1$ and $-1$ meaning right and left handed, respectively.}
\end{figure}

To put our EH observations into the context of the Cherenkov mechanism, we plot the kinetic (orange and pink from WHAMP~\cite{ronnmark1982}) and cold (blue) whistler dispersion relation ($k_\perp=0$) for one group of slow EHs ($v_{EH}\approx v_{Ae}/16$) with $T_\perp/T_\parallel=1.0$ in Fig.~\ref{fig:cherenkov}a, and for one group of fast EHs ($v_{EH}\approx v_{Ae}/4$) with $T_\perp/T_\parallel=0.3$ in Fig.~\ref{fig:cherenkov}b. We define and plot $\omega_{EH}=\pi/t_{pp}$, where $t_{pp}$ is the peak-to-peak time of $\delta E_\parallel$, and $k_{EH}=\omega_{EH}/v_{EH}$, color-coding $\delta B_\perp$. The Cherenkov resonance condition is for a given EH manifested in the plots as the intersection of $\omega_r(k_\parallel)$ with the straight line passing through the origin and the point $(k_{EH},\omega_{EH})$. The slope of this line corresponds to $v_{EH}$, meaning faster EHs excite whistlers with smaller $k_\parallel$. The shaded regions contain EH velocities between $\text{max}(v_{EH})$ and $\text{min}(v_{EH})$ for the two groups.

 For the slow EHs (Fig.~\ref{fig:cherenkov}a), these intersections occur at $k_\parallel d_e\gg1$. However, for the fast EHs (Fig.~\ref{fig:cherenkov}b) we find that the EHs can excite whistlers in the wavenumber range $2.3 \leq k_\parallel d_e \leq 4.7$. This interval is marked by the blue vertical lines at the intersection for the fastest and slowest EHs.
 We note that there is an additional permitted region for small $k_\parallel d_e\ll1$, which was observed in Ref.~\cite{goldman2014}. For the observed EHs however, $k_\parallel\approx k_{EH}$, which is consistent with waves in the larger $k_\parallel$ interval. 

For the permitted waves in the larger $k_\parallel$ interval, $\gamma$ is large and negative. The resonant whistlers are thus strongly damped, providing a possible explanation to why $\delta\mathbf{B}_{\text{Res},\perp}$ is typically confined within the EHs.
Note that we are investigating the classic Cherenkov mechanism, where waves are excited by a propagating charge acting as an antenna \cite{singh2001,compernolle2008}, not by kinetic Landau resonance. This is why the growth from the Cherenkov mechanism does not appear in Fig.~\ref{fig:cherenkov}.

Extending the dispersion relation in Fig.~\ref{fig:cherenkov}b to include $k_\perp>0$ yields the surface in Fig.~\ref{fig:cherenkov}c and \ref{fig:cherenkov}d, showing the relative damping $\gamma/\omega_r$ and ellipticity respectively. By including $k_\perp>0$, the resonant waves go from being points on a curve, to contours on a surface. The blue contours in Figs.~\ref{fig:cherenkov}c,d show the waves that can be excited by the fastest and slowest EHs in Fig.~\ref{fig:cherenkov}b, meaning the other EHs in Fig.~\ref{fig:cherenkov}b can excite whistlers between these contours. From observations we have ellipticity values close to 1, consistent with the permitted $k_{\perp}\lesssim k_{\parallel}$ region in Fig.~\ref{fig:cherenkov}d. 

Additionally, the fact that we observe a strong $v_{EH}/v_{Ae}$ dependence of $\delta B_\perp$ (Fig.~\ref{fig:stats}) is explained by the $(v_{EH}/v_{Ae})^4$ dependence of the secular whistler growth. $v_{EH}/v_{Ae}$ is 4 times larger for the EHs in Fig.~\ref{fig:cherenkov}b than for those in Fig.~\ref{fig:cherenkov}a, meaning they grow $\sim250$ times faster.
This explains why significant $\delta\mathbf{B}_{\text{Res},\perp}$ is observed only for the fast EHs as was found in Fig.~\ref{fig:fit}.

As an example we consider the EH with the tail-like $\delta\mathbf{B}_\perp$ shown in Figs.~\ref{fig:overview}g,j. This EH is located at the point $k_{EH}d_e=2.0$, $\omega_{EH}/\omega_{ce}=0.55$ in Fig.~\ref{fig:cherenkov}b, and its velocity $v_{EH}=0.28v_{Ae}$ corresponds to the black line. From the Cherenkov resonance condition we expect the emitted whistler to have $\omega/\Omega_{ce}=0.73$ and $k_\parallel d_e=2.7$. The EH is observed by all four MMS spacecraft and we apply a generalized four-spacecraft version of the method discussed in Ref.~\onlinecite{graham2016} on $\delta\mathbf{B}_\perp$ to determine $\omega/\Omega_{ce}=0.76$ and $k_\parallel d_e=3.2$. This point is marked in Fig.~\ref{fig:cherenkov}b with a black cross. The predicted damping for the observed wave is $\gamma\approx-0.25\Omega_{ce}$, qualitatively consistent with the strong decay seen in Fig.~\ref{fig:overview}j. Taking the observed $k_\perp d_e=0.53$ into account in Figs.~\ref{fig:cherenkov}c,d, the black contour corresponds to the Cherenkov resonant waves, and we see that the observed wave (black cross) is still close to the modes predicted by the Cherenkov mechanism.
We thus conclude that the Cherenkov mechanism is in good agreement with observations, and is likely the source of $\delta\mathbf{B}_{\text{Res},\perp}$.

\paragraph*{Conclusions.}
In summary, we report MMS observations of electron holes (EHs) with magnetic field signatures consisting of monopolar $\delta\mathbf{B}_\parallel$ and right hand polarized $\delta\mathbf{B}_\perp$. Typically, $\delta\mathbf{B}_\perp$ is confined within the EH and only one wave period is observed. In rare cases however, multiple periods can be observed extending outside the EH while rapidly decaying. The frequency of $\delta\mathbf{B}_\perp$ is below $\Omega_{ce}$. Using spacecraft timing we calculate $v_{EH}$ and $\Phi_m$, finding $\delta\mathbf{B}_\perp$ to correlate with both parameters. 
We are able to calculate the magnetic field generated by $\delta\mathbf{E}\times\mathbf{B}_0$ drifting electrons, $\delta\mathbf{B}_d$, in a few cases, concluding that this mechanism is responsible for the observed $\delta\mathbf{B}_\parallel$, and that $\delta\mathbf{B}_L\ll\delta\mathbf{B}_d$, where $\delta\mathbf{B}_L$ is the Lorentz transform of the EHs electric field, in the observed parameter range. For slow EHs ($v_{EH}/v_{Ae}\lesssim 0.1$) $\delta\mathbf{B}_\perp\approx\delta\mathbf{B}_{d\perp}$, whereas an additional $\delta\mathbf{B}_\perp$ source is required for faster EHs.
We show that this additional field is consistent with whistler waves generated by EHs via the classic Cherenkov mechanism (not Landau resonance). This is supported by the right-hand polarization and $\omega<\Omega_{ce}$, and the fact that significant $\delta\mathbf{B}_\perp$ is observed for EHs with speeds approaching $v_{Ae}/2$. The kinetic whistler dispersion relation shows that there is significant damping for the wavenumbers predicted from the Cherenkov mechanism, which suggests that mainly a near-field signal will be excited. This is consistent with our observation of $\delta\mathbf{B}_\perp$ being localized to the EH itself. 

Using multi-spacecraft MMS observations we can for the first time quantify individual contributions to $\delta\mathbf{B}$ of EHs. We report the first observational evidence of EHs Cherenkov radiating whistler waves, though the waves tend to be localized within the EHs rather than freely propagating.

\begin{acknowledgments}
\paragraph{Acknowledgements.}
We thank the entire MMS team and instrument PIs for data access and support. MMS data are available at https://lasp.colorado.edu/mms/sdc/public. This work is supported by the Swedish National Space Board, grant 128/17, the Swedish Research Council, grant 2016-05507. French involvement (SCM instruments) on MMS mission is supported by CNES and CNRS.
\end{acknowledgments}

\bibliography{biblio.bib}

\end{document}